\documentclass[aps,showpacs,superscriptaddress]{revtex4}

\usepackage{graphicx}
\usepackage{amsmath,amssymb,latexsym}
\usepackage{color}
\bibstyle{apsrev.bib}

\begin{document}
\def\be{\begin{equation}}
\def\ee{\end{equation}}

\def\bfi{\begin{figure}}
\def\efi{\end{figure}}
\def\bea{\begin{eqnarray}}
\def\eea{\end{eqnarray}}

\title{Roughening of an interface in a system with surface or bulk disorder.}

\author{Federico Corberi}
\affiliation {Dipartimento di Fisica ``E.~R. Caianiello'', and INFN, Gruppo Collegato di Salerno, and CNISM, Unit\`a di Salerno,Universit\`a  di Salerno, 
via Giovanni Paolo II 132, 84084 Fisciano (SA), Italy.}

\author{Eugenio Lippiello}
\affiliation{Dipartimento di Scienze Ambientali, Seconda Universit\`a di Napoli,
Via Vivaldi, Caserta, Italy.}

\author{Marco Zannetti}
\affiliation {Dipartimento di Fisica ``E.~R. Caianiello'', and INFN, Gruppo Collegato di Salerno, and CNISM, Unit\`a di Salerno,Universit\`a  di Salerno, via Giovanni Paolo II 132, 84084 Fisciano (SA), Italy.}

\date{\today}

\begin{abstract}
We study numerically the roughening properties of an interface in a
two-dimensional Ising model with either random bonds or random fields,
which are representative of universality classes where disorder acts only on the interface
or also away from it, in the bulk. The dynamical structure factor shows a rich 
crossover pattern from the form of a pure system at large wavevectors $k$, 
to a different behavior, typical of the kind of disorder, at smaller $k$'s. 
For the random field
model a second crossover is observed from the typical behavior of a system
where disorder is only effective on the surface, as the random bond model, to the 
truly large scale behavior, where bulk-disorder is important, 
that is observed at the smallest wavevectors.

\end{abstract}


\maketitle

\section{Introduction}

Interface roughening is a widely observed feature 
occurring in many fields of science ranging from physics to engineering
and biology, including crystal growth, viscous flow in porous media 
and many other phenomena \cite{15pleim}.

Although interfaces may roughen also in the absence of disorder,
a great deal of interest has been recently payed to 
those systems whose properties are determined by the presence
of impurities or other sources of quenched randomness.
Typical examples are found in high-Tc 
superconductors \cite{da45a51pleim},
and disordered magnets \cite{refscrp}.
In these systems the properties of the interfaces are mainly due to the 
interplay between their elasticity and the localization effects due to the
impurities. 
In a superconductor, the bending of magnetic flux lines
is due to the competition between defects pinning (grain boundaries, oxygen impurities etc.) and the 
bending energy \cite{4kard}. 
Similarly, the geometry of domains walls in disordered ferromagnets 
originates from the balance between the energy gain associated 
to interface deformations caused by random impurities and the surface tension cost.
This mechanism produces peculiar equilibrium and non-equilibrium properties \cite{5kard}.

A great advance in roughening theory
has been achieved by the recognition that a scaling symmetry, analogous
to the one at work in ordinary critical phenomena, underlies the properties  
of interfaces \cite{familvick}. 
According to this hypothesis, when roughness develops in an initially 
smooth interface at time $t=0$, 
the evolution of the structure factor [see around Eq. (\ref{corrg}) for 
a precise definition of this quantity] takes the form  
\be
S(k,t)\simeq k^{-(1+2\zeta)}f[k\xi (t)],
\label{s_clean0}
\ee
where $k$ is the wavevector, $\zeta $ is the roughness exponent, $\xi (t)$ has
the physical dimensions of a length and $f$ is a scaling function. 
The behavior of these quantities and, in particular, of the roughness
exponent, are expected to exhibit universal properties.
One of  the relevant features determining the universality classes is the presence of quenched disorder. 
Indeed, in many cases it
is found to modify the properties of the interfaces with
several consequences on the materials properties, as it is observed, for instance,
in the growth of a magnetic phase \cite{refscrp2}.

Not only the presence of quenched randomness is relevant, but also its nature.
An important distinction exists \cite{natterman} between systems where disorder 
acts only at the interfaces (interface disorder) and others where it is effective 
also in the regions away from them (bulk disorder). 
Systems belonging to these two classes are all expected to obey Eq. (\ref{s_clean0})
but with different roughening exponents and properties of $\xi$ and $f$. 

In this paper we investigate numerically the behavior of a one-dimensional
interface evolving in a two-dimensional random medium. The random bond
and the random field Ising models are taken as prototypical of systems with interface disorder
and with bulk disorder, respectively.
By computing the structure factor we show the existence of a 
crossover from pure to interface-disorder behavior or from pure to
bulk-disorder behavior as the  wave vector is changed.
Usually \cite{refscrp2}, to randomness is associated a characteristic length $\lambda$, and
the crossover from pure to disorder-dominated behavior takes place as the
observation length scale grows from smaller to larger than $\lambda$.
Indeed, this is the pattern observed in the interface disorder case, i.e. with random
bonds, where the structure factor is pure-like in the wave vector range
$k\gg \lambda ^{-1}$ and departs from pure behavior for $k \ll \lambda ^{-1}$.
Instead, the crossover in the bulk disorder case due to the random fields
is novel and more interesting. In fact, the data can be accounted for by introducing
an additional characteristic length scale $\Lambda > \lambda$, such that
the crossover pattern displays pure behavior for $k\gg \lambda ^{-1}$,
interface-disorder behavior for  $k$ in between $\Lambda ^{-1}$ and $\lambda ^{-1}$,
and bulk-disorder behavior for $k \ll \Lambda ^{-1}$. In particular, in the intermediate regime
the roughness exponent takes the value 
($\zeta =2/3$ \cite{natterman,dueterzi}) expected for a system with interface-disorder.
Truly large-scale properties are only displayed on lengthscales
larger than $\Lambda $, where 
the asymptotic value of the 
roughness exponent takes the value expected in a system with bulk disorder ($\zeta =1$ \cite{natterman}).

The paper is organized in five sections: In Sec. \ref{themodel} 
we introduce the random bond and the random fields Ising models.
In Sec. \ref{theclean} we consider, preliminarly, the 
pure case in order to have a reliable benchmark to compare with,
particularly for what concerns the relevance of finite size effects.
Sec. \ref{thedisorder} is devoted to the study of the disordered cases 
and to the discussion of what is observed. In Sec. \ref{theconclusion}
we summarize the main results.
An appendix contains a further discussion of the finite-size effects of
the model.

\section{The model} \label{themodel}

We consider a ferromagnetic system 
described by the Hamiltonian
\be
{\cal H}(\{S_i\})=-\sum _{\langle ij\rangle}J_{ij}S_iS_j+\sum _i H_iS_i,
\label{isham}
\ee
where $S_i=\pm 1$ are Ising spin variables defined on a two-dimensional
$L\times L$ square lattice and $\langle ij\rangle$ are two nearest 
neighbor sites. 
We denote by $x$ ($x=1,2,\dots,L$) and $z$ 
the coordinates of a site $i$ of the lattice 
along the horizontal and vertical direction, respectively.    

We will study two types of quenched disorder:

i) Random bonds Ising model (RBIM): In this case $H_i\equiv 0$ and the coupling constants $J_{ij}$
are independent random numbers extracted from a flat distribution in 
$[J_0-\delta,J_0+\delta]$, with $\delta <1$ in order to keep the interactions
ferromagnetic and to avoid frustration effects.

ii) Random fields Ising model (RFIM): In this case the ferromagnetic bonds
are fixed $J_{ij}\equiv J_0$, while the external field at each site $H_i=\pm \delta$ is
uncorrelated and sampled from a symmetric bimodal distribution.

\begin{figure}[!t]
\begin{center}
\includegraphics[width=0.65\columnwidth]{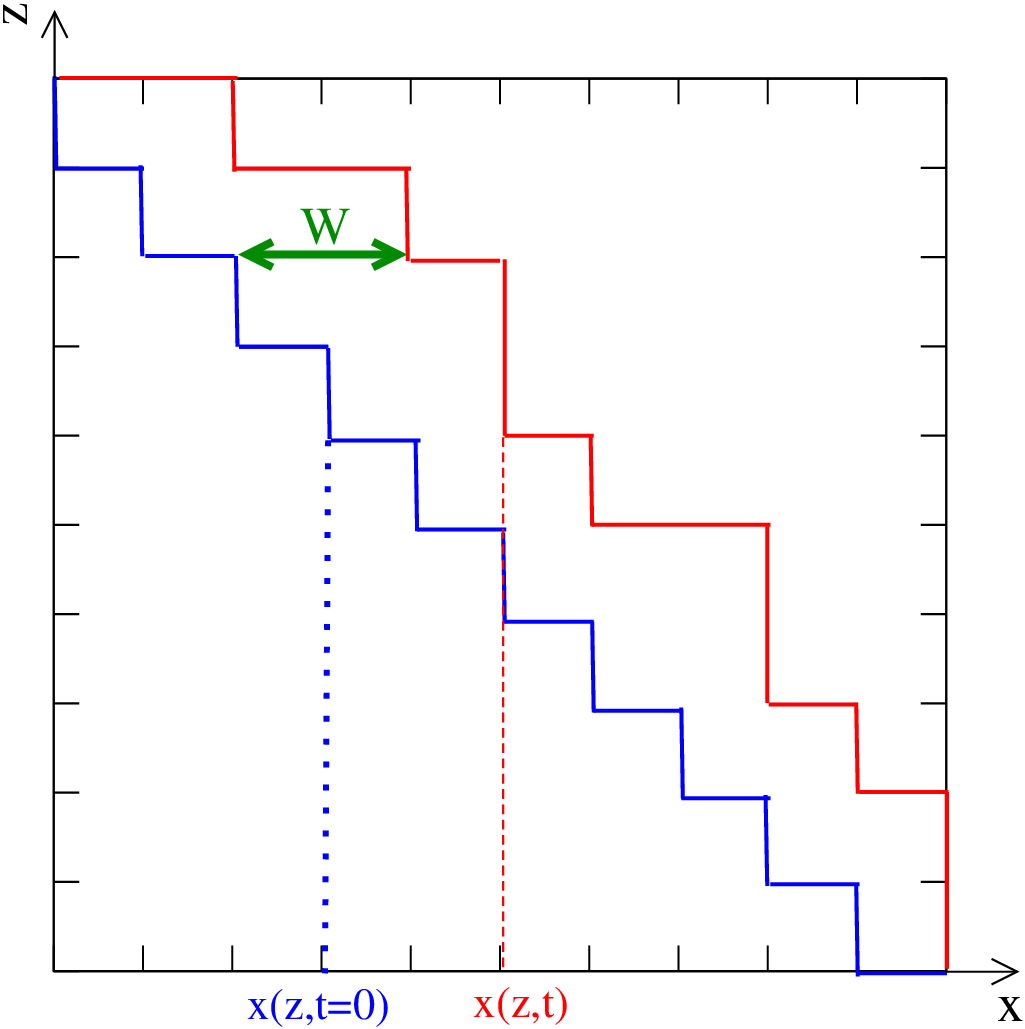}
\end{center}
\caption{The position of the interface at time $t=0$ (blue curve on the left)
and its evolution at a later time $t$ (red curve on the right) are pictorially sketched.
$W(t,\epsilon)$ is of the order of magnitude of average horizontal distance between the two profiles.}
\label{fig_sketh}
\end{figure}

We prepare the system at time $t=0$ by seeding an interface ${\cal I}$ along the diagonal, 
i.e. we set spins $S_i=+1$ in one of the two halves of the system with $z\ge L-x$ and 
$S_i=-1$ in the other, as pictorially shown in Fig. \ref{fig_sketh}. 
Then, moving horizontally at a given value 
of $z$, upon increasing $x$ one encounters only one couple of neighboring 
antiparallel spins $S_iS_j$ and
we define the (horizontal) position $x(z,t=0)$ of the interface 
as the coordinate $x$ of the first non-aligned spin $S_i$.  

A dynamics is then introduced from time $t=0$ onwards by flipping
single spins according to Glauber transition rates at temperature $T$
\be
w(S_i\to -S_i)=\frac{1}{2}\left [1-S_i\tanh \left (\frac{H^W_i+H_i}{T}\right )\right ]
\ee
where 
\be
H^W_i=\sum _{j\in  nn(i)}J_{ij}S_j
\ee
is the local Weiss field (the sum runs over the nearest neighbours $j$ of $i$).

In order to keep a single spanning interface stable at all times 
we use anti-periodic boundary conditions and set $J_0\to \infty$;
this can be implemented practically by forbidding flips of spin 
away from the interface. Notice that with this choice the equilibrium-state
of the Ising model (\ref{isham}) is ferromagnetic at any finite temperature.
Furthermore, in this limit the model contains a single
parameter $\epsilon =\delta/T$, so in the following we will fix $T=0.01$ and
let $\delta $ vary. Since we are interested in the limit $L \to \infty$
of an infinite system we will set $L$ to the largest numerically tractable
value. A discussion of unavoidable finite-size effects will be given 
in the following sections and in the Appendix.

Before proceeding, let us stress the fact that in both model i) and ii) 
there is no neat drive on the motion of the domain wall,
because the average value $\overline {H_i}$ of $H_i$ vanishes.
This is at variance with the frequently studied 
problem \cite{sugref1,sugref2} where the depinning of the interface is obtained 
by raising $\overline {H_i}$ above a certain threshold value.

The main dynamical feature of the process is the 
kinetic roughening of the interface which can be studied 
by computing the structure factor $S(k,t,\epsilon)$, namely
the Fourier transform (with respect to $r$) of the surface correlation function
\be
G(r,t,\epsilon)= \frac{1}{L} \int_0^L \langle u (z,t)u(z+r,t)\rangle _\epsilon dz,
\label{corrg}
\ee
where $u(z,t)\equiv x(z,t)-x(z,0)$ is  
the horizontal displacement of the interface, 
namely the distance moved from its initial position, and $\langle \cdots \rangle _\epsilon$
is a statistical average in the presence of disorder, if present. 

The interface roughness $W$ is defined by the autocorrelation function
\be
W^2(t,\epsilon)=\frac{1}{L} \int_0^L \langle u^2 (z,t)\rangle _\epsilon dz=
G(0,t,\epsilon)
\ee
which is obtained from the structure factor as
\be
W^2(t,\epsilon)=\int \frac{dk}{2\pi}\,S(k,t,\epsilon).
\label{wfroms}
\ee

\section{Pure system} \label{theclean}

Before studying the effects of quenched disorder it is useful to 
consider preliminarly the well-known behavior of the structure factor in the pure case.
This will serve as a benchmark to tune the methodology
and to discuss the presence of finite-size effects. 

For a pure system ($J_{ij}\equiv J_0,H_i\equiv 0$) of infinite size,
continuum theories \cite{upton} predict the scaling form (\ref{s_clean0})
for $t$ sufficiently large, $\zeta =1/2$,
and a scaling function satisfying the limiting behaviors
\be
f(x)= \left \{ 
\begin{array}{ll} const.\,, & \mbox{for}\,\,\, x\gg 1\\
x^{1+2\zeta}\,\,, & \mbox{for}\,\,\, x\ll 1
\end{array}
\right . .
\label{s_clean}
\ee    
This is correct also
on a lattice for wavevectors much smaller than the
inverse lattice spacing $a^{-1}$. In order to make the extension 
to all $k$ values the replacement $k\to q=\frac{1}{a}\sqrt{2[1-\cos(ka)]}$
must be made in Eqs.~(\ref{s_clean0},\ref{s_clean}). This monotonic 
transformation amounts only to a small deformation at the largest wavevectors
and extends the validity of Eq. (\ref{s_clean0}) to the large $k$ domain. Given the very similar
role played by $k$ and $q$, we will use the term 
{\it wavevector } also for $q$ (with a minor abuse of language ). 

Notice that Eq. (\ref{s_clean0}) holds in an infinite system with $L=\infty$.
For all practical purposes the condition $\xi (t)\ll L$ is sufficient to guarantee
the infinite-system behavior. Since $\xi (t)$ is a growing length, given a certain 
system size $L$, this condition is fulfilled for times $t\ll t_L$, where $t_L$ is such that
$\xi (t_L)\sim L$. 
As we will discuss further below, in numerical simulations
it is important to reach very long times in order to have a good estimate
of the asymptotic properties we are interested in, but this implies that finite-size
effects may be present. In order to keep these corrections under control, the scaling form (\ref{s_clean0}) 
must be upgraded to take into account the system's finite size
\be
S(k,t,L)=q^{-(1+2\zeta)}g\left [q\xi (t),qL\right ],
\label{s_clean2}
\ee
where we have replaced $k$ with $q$ as discussed above, and
$g(x,y)$ is such that no finite size effects are present for $y=qL\gg 1$, i.e.
\be
g(x,y)\simeq f(x)\,, \,\,\,  \mbox{for}\,\,\, y\gg 1,
\label{s_cleang1}
\ee
with $f$ the same function of Eqs.  (\ref{s_clean0},\ref{s_clean}),   
whereas for $x\gg 1$ (i.e. for large times) $S(k,t,L)$ converges to the equilibrium 
value $S_{eq}(k,L)$ which explicitly depends on $y$ and determines, 
using Eq. (\ref{wfroms}), the saturation to the 
equilibrium value of the roughness 
\be
W^2(t=\infty,L)=L^{2\zeta},
\label{s_cleang2}
\ee
thus earning $\zeta $ the name {\it roughness exponent}. 

\begin{figure}[!t]
\begin{center}
\includegraphics[width=0.95\columnwidth]{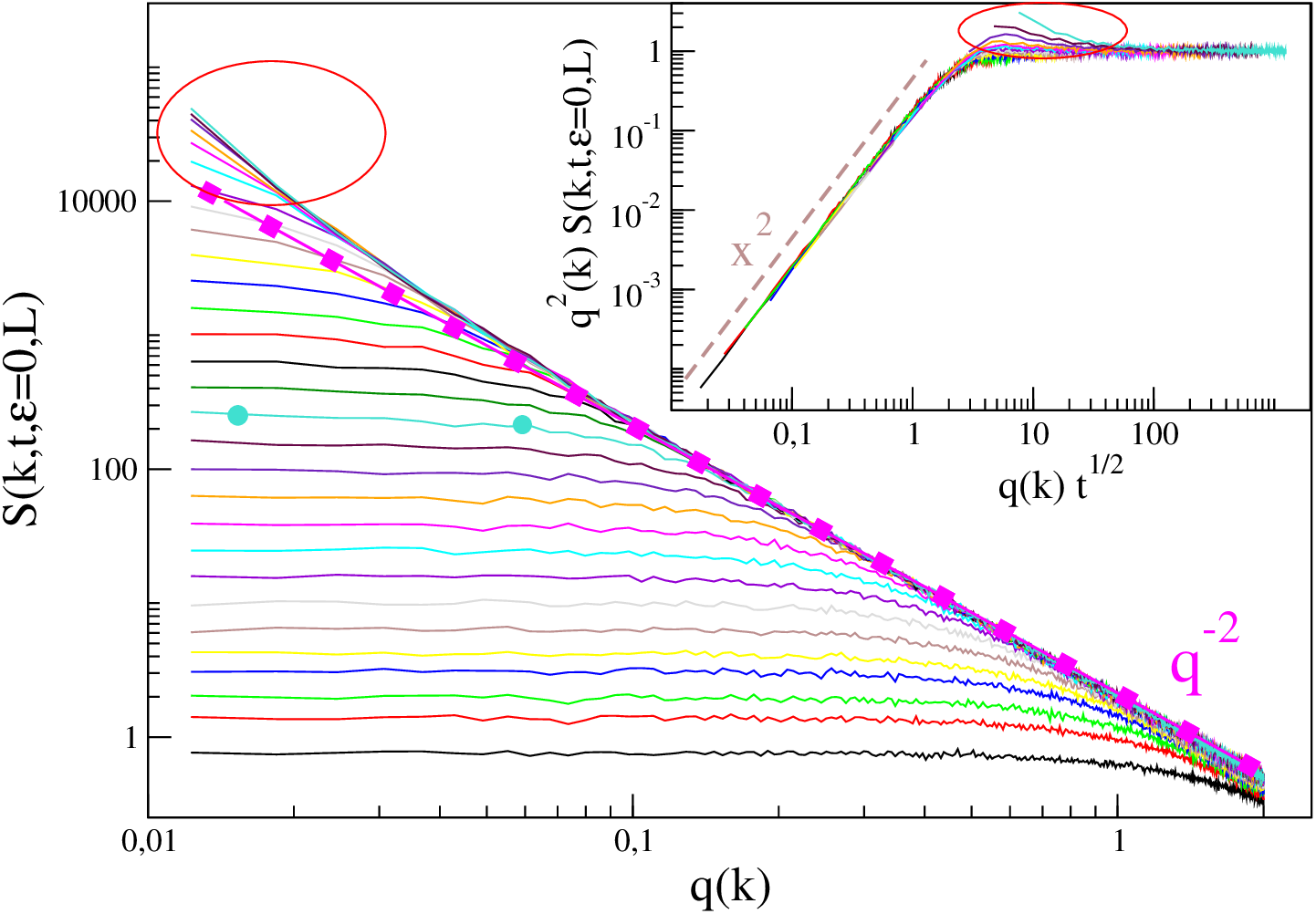}
\end{center}
\caption{$S(k,t,\epsilon=0,L)$ is plotted against $q$ for the pure case.
Different curves correspond to exponentially increasing times 
($t=2,3,5,7,11,18,29,46,73,118,189,304,490,789,1269,$ $2044,3291,5299,8532,13739,$
$22123,35623,57362,92368,148736,239503,385663,621017,10^6$
from bottom to top).
The bold-dotted magenta line
is the power-law $q^{-2}$ expected for
an infinite system in the limit $t\to \infty$. The regions encircled by a red
ellipse is the one where finite-size effects show-up.
In the insets the data collapse obtained by plotting $q^2S(k,t,\epsilon=0,L)$ 
against $qt^{1/2}$ is shown. The brown-dashed line is the power-law
$x^2$ expected for small $x$ according to Eq. (\ref{s_clean}).}
\label{fig_clean}
\end{figure}

Let us now illustrate how this scaling scenario works. 
In Fig.~\ref{fig_clean}, the structure factor $S(k,t,L)$ for the pure system is plotted 
against $q$ for a system of size $L=1024$ and several times $t$.
As expected from 
Eqs.~(\ref{s_clean0},\ref{s_clean}), in the absence of finite-size effects  
(i.e. for sufficiently small times $t\le 92368$)
these curves obey the power law $S\sim q^{-2}$ for wavevectors
larger than a typical value that we identify with $\xi ^{-1}(t)$, and then
bend over to a constant for $q\ll \xi ^{-1}(t)$. The length $\xi (t)$
can be operatively extracted by asking for the best data-collapse
of the scaling function $q^2S$ at different times as a function of
$q\xi(t)$, yielding the well known result $\xi (t)\sim t^{1/2}$. 
The data collapse obtained by plotting $q^2S$ against $qt^{1/2}$
is shown in the inset of Fig.~\ref{fig_clean}.

The curves for $t\ge 148736$ (largest five times) in Fig. \ref{fig_clean} correspond
to observation times large enough to reach saturation at $f(x)=const.$ in Eq. (\ref{s_clean}) in the
whole available range of wavevectors. Then one should observe 
$S\propto q^{-2}$ for all $q$. 
Instead, these curves bend upward at small wavevectors (for $q\lesssim 0.03$,
in a region that has been indicated by an ellipse in Fig. \ref{fig_clean})) 
and depart from the behavior expected in an infinite system. The same effect is observed
in the inset of the figure, where data collapse fails at the largest times
for small $qt^{1/2}$.
This is a finite size effect, since the onset wavevector becomes
smaller by increasing the system size $L$. Further evidence for this is given in the
Appendix.
According to Eq. (\ref{s_clean2}), this means that for such long times $\xi $ is so
large to make relevant  the second entry of the scaling function $g$, spoiling 
in this way the 
infinite system behavior (\ref{s_clean0},\ref{s_clean}). 

Recalling Eq. (\ref{s_clean2}) it is clear that the roughness 
exponent can be obtained by fitting the slope $-1+2\zeta $ of the plot of $S$ 
versus $q$ in the regime where the scaling function $g$ is constant,
namely when the wave vector condition
\be
q\gg \max \left[\xi ^{-1},L^{-1}\right]
\label{twocond}
\ee
is met. In the pure case this amounts to consider the linear envelope 
of the various curves (for different times) of $S$ versus $q$ in the double-logarithmic plot
of Fig. \ref{fig_clean}.
Considering, as an example, the curve at the intermediate time $t=789$ (turquoise with
circles) it is 
clear that the linear envelope is
not followed for $q\lesssim 0.2$ because the curve flattens as due to the fact that 
$q\gg \xi ^{-1}$ is no longer obeyed (one can easily be convinced that
$\max \left[\xi ^{-1},L^{-1}\right]=\xi ^{-1}$ in this case  by the fact that 
the typical wavevector where $S$ starts flattening is a decreasing function of time).  
Operatively, we define the region $q\gg \xi ^{-1}$
as the one where the data at a certain time superimpose to those at the immediate earlier times.
The situation changes for very large times ($t\ge 35623$, say). For such long times one 
sees that the linear envelope is no longer followed 
for $q\lesssim 0.05$. In this case, at variance with what observed for intermediate times
(discussion above),
there is an initial upward bending of the curves in a region marked by the red ellipse 
(before flattening to a finite value for $q\to 0$). In this case one has 
$\max \left[\xi ^{-1},L^{-1}\right]=L^{-1}$, as it can be recognized by the fact that
the curves for $S$ at different times all depart from the linear envelope around the 
time-independent value $q\simeq 0.05$ (since, differently from $\xi ^{-1}$, 
$L^{-1}$ is time independent).
In conclusion, for any time $t$ one can practically define the $q$-region where the condition 
(\ref{twocond}) is met as that where the data for $S$ superimpose to those at the earlier time
and $q\ge 0.05$. We will see in the following that an analogous procedure can be adopted for the cases
with disorder.

\section{Quenched disorder} \label{thedisorder}

As anticipated in the Introduction and as discussed in Ref.~\cite{refscrp2},
when quenched disorder parametrized by $\epsilon$ is added to a system its 
effects are usually relevant on lengthscales larger than
a certain characteristic length $\lambda (\epsilon)$.
For instance, it is clear that introducing a small density $\rho$ of diluted impurities 
in a pure system cannot change the properties \cite{lengthlambdadil} 
on lengthscales much smaller
than the typical distance $\simeq \rho ^{-1/d}$ between such impurities ($d$ is
the spatial dimension), simply because within such small distances the probability to
find an impurity is negligible and the system is basically undistinguishable
from the pure one. The same situation is found also in models,
as the disordered Ising magnets considered here,
or bonds, where {\it impurities} are not diluted but 
distributed
all over the system, although the physical interpretation of this length
might not be straightforward.
The presence and the properties of this length in the RBIM and RFIM
were discussed in \cite{lengthlambdarb} and ~\cite{lengthlambdarf}
where it was shown that $\lambda $ is a decreasing function of $\epsilon$
with the property
\be 
\lim _{\epsilon \to 0} \lambda (\epsilon)=\infty.
\label{property}
\ee 

Since disorder is effective over distances larger than $\lambda$, 
in an infinite system the pure form~(\ref{s_clean}) of $S$, with
$\zeta =1/2$, is expected to be obeyed for 
$q\gg \lambda ^{-1}(\epsilon)$ even in the presence of disorder.
In order to check this and other issues that will be addressed below,
we have performed a 
series of simulations of the RBIM and of the RFIM.
For the former we have considered a system of size $L=512$, while for the
latter we set $L=1024$ for $\epsilon \le 0.5$, $L=512$ for 
$\epsilon = 0.6$ and $L=256$ for $\epsilon = 0.8$. 
Smaller system sizes are used for the largest values of $\epsilon $ since
the dynamics is slower.

\begin{figure}[!t]
\begin{center}
\includegraphics[width=0.95\columnwidth]{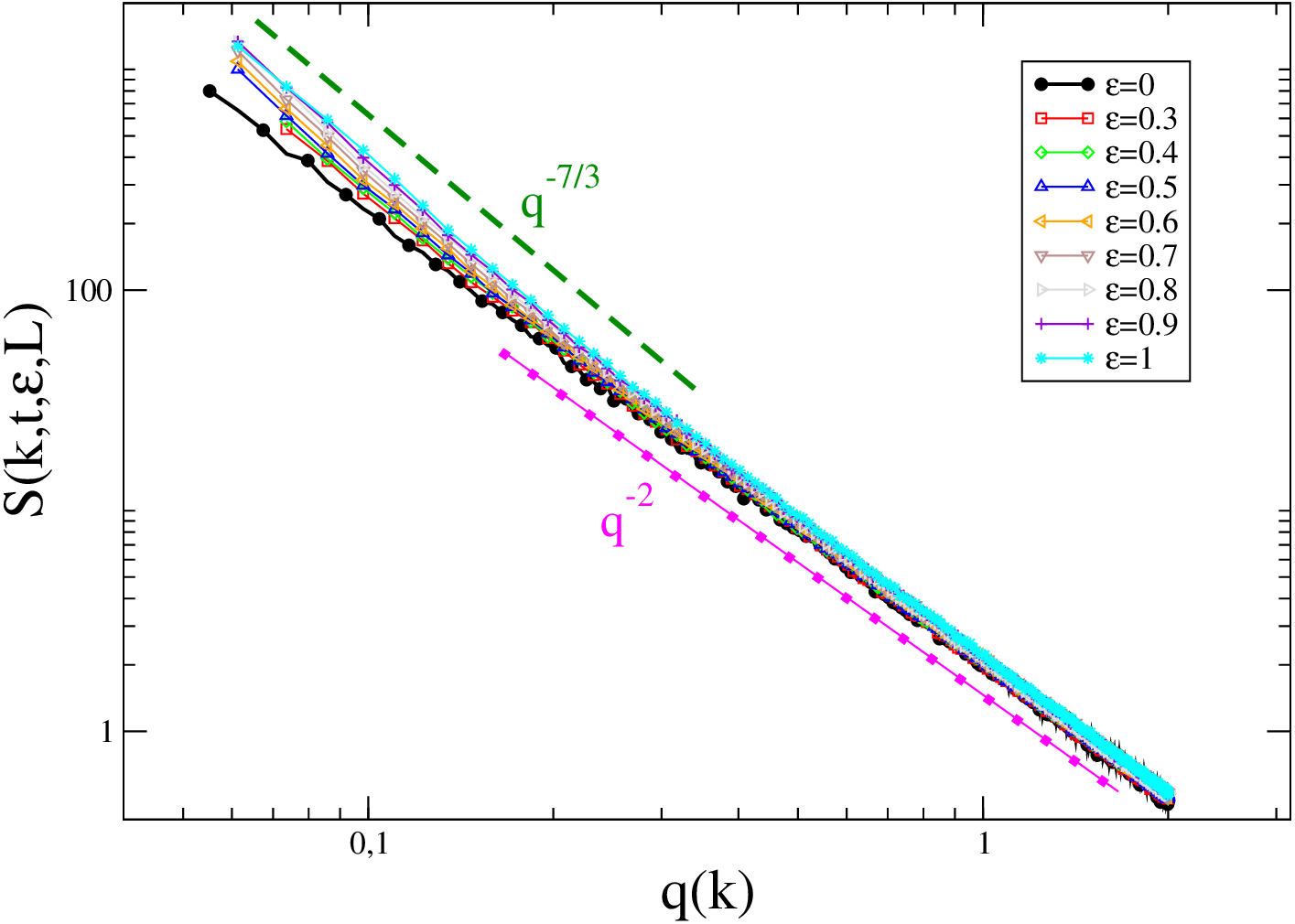}
\end{center}
\caption{$S(k,t,\epsilon,L)$ is plotted against $q$ for the RBIM at
$t=10^7$ ($t=10^6$ for $\epsilon =0$).}
\label{fig_rb}
\end{figure}

\begin{figure}[!t]
\begin{center}
\includegraphics[width=0.95\columnwidth]{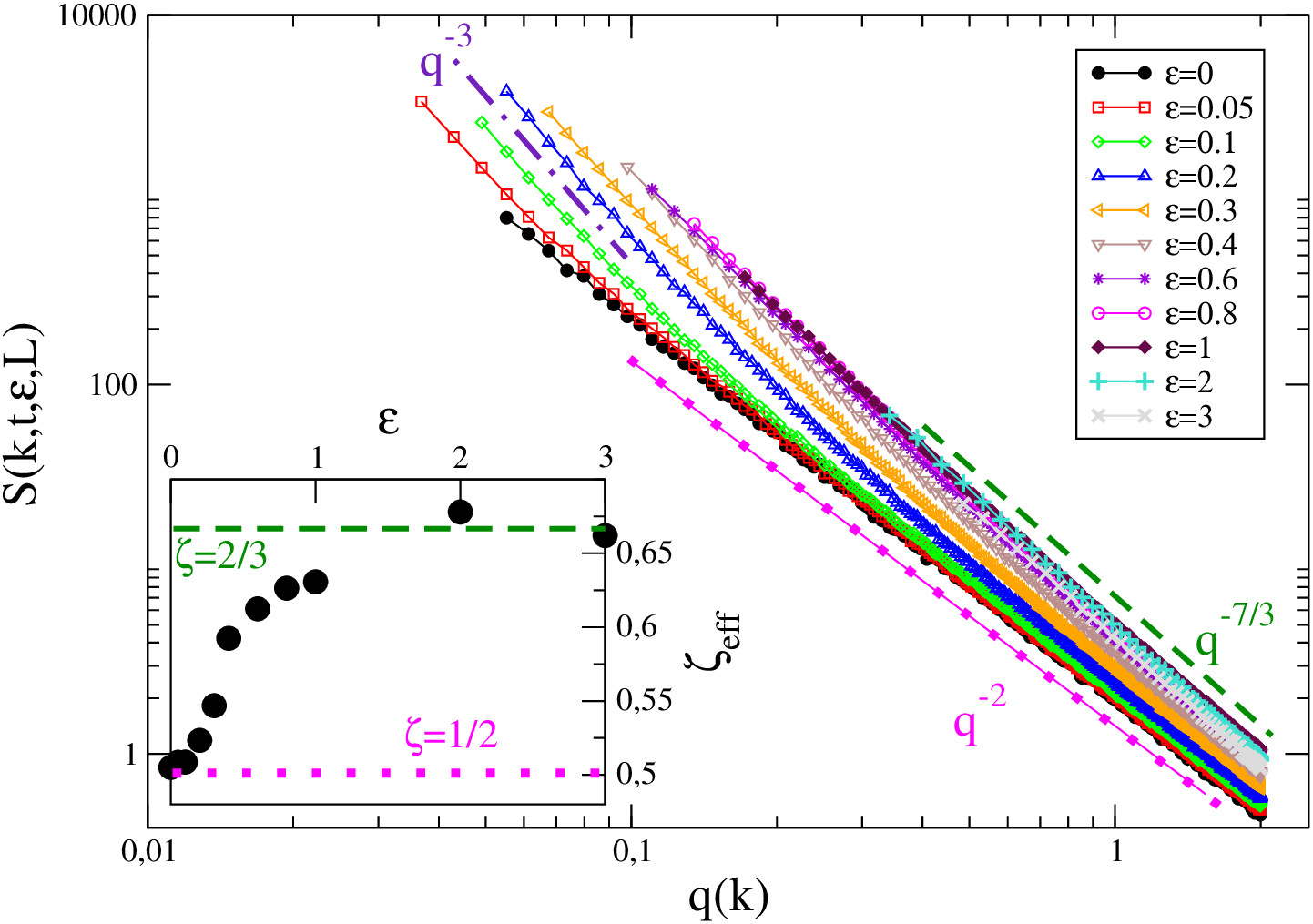}
\end{center}
\caption{$S(k,t,\epsilon,L)$ is plotted against $q$ for the RFIM at
$t=10^6$ ($t=10^7$ for $\epsilon \ge 0.7$). In the inset the {\it effective exponent}
$\zeta $ obtained by fitting the curves for $q\ge 1$ is plotted against $\epsilon$.}
\label{fig_rf}
\end{figure}

The structure factor for the RBIM is shown 
in Fig.~\ref{fig_rb} at time $t=10^7$.
Here it is observed that $S$ behaves as in the pure
system down to wavevectors of order $q\simeq 0.2-0.3$. In particular, 
in this range, the curves are independent of $\epsilon$
and behave as $q^{-2}$.
For smaller values of the wavevector
the curves for the disordered system deviate upward 
from the $q^{-2}$ pure behavior, signaling
a crossover to a region where disorder is relevant.
Despite the smoothness of the crossover, it can be observed that
the curves for large $\epsilon$ depart from the pure behavior at larger wave vectors, 
and in a more pronounced way.
This is because the crossover occurs when 
$q^{-1}\simeq \lambda (\epsilon)$ and, as discussed above 
(recall Eq. (\ref{property})), the latter is a 
decreasing function of $\epsilon$. 
Notice also that for the largest values of $\epsilon $ 
a small departure from the pure-like behavior can be detected even for the largest
wavevectors: starting from
$\epsilon =0.9$ the curves appear slightly displaced vertically.
This signals the fact that, for these large values of the disorder strength,
$\lambda ^{-1}$ is already of the order of the largest wave vectors and the effects
of an incipient crossover can already be detected.  
The discussion of
$S$ in this range of $q$ where disorder is effective is postponed to
Sec. \ref{ranbo}. 

Limited to the large-$q$ sector, a similar behavior of $S$ 
is observed also in Fig.~\ref{fig_rf} for the model with random fields.
Here we compare the structure factor computed at the longest simulated time
($t=10^6$ for $\epsilon \le  0.6$ or $t=10^7$ for $\epsilon =0.8$)
for different values of $\epsilon$.
When the strength of the disorder is moderate, for $\epsilon \lesssim 0.1-0.2$,
the curves for large $q$ superimpose on the one for $\epsilon =0$. 
However, for smaller wave vectors, they bend upwards very evidently.
For larger values of the disorder strength, for $\epsilon \gtrsim 0.3$, the curves
start to depart from the pure case already at the largest wave vectors considered.
In the same range of values of $\epsilon $ the departure is much larger than in
the RBIM and, for the largest value of $\epsilon$ a slope somewhat larger
than $q^2$ is observed, as will be further commented in Sec. \ref{ranfi}.

Up to this point we have checked that the pure-like behavior is obeyed also in the
presence of a sufficiently small disorder in the large wavevector sector.
Since time is very large in Figs. \ref{fig_rb},\ref{fig_rf}, in this large-$q$ region
one always has
$q\xi (t)\gg 1$ and no finite-size effects. 
In the following we will discuss how the properties of the roughness change
for smaller $q$ where the presence of quenched randomness becomes relevant
(upward bending of the curves).
We still want to focus on the region where condition (\ref{twocond}) is satisfied because, according to 
the discussion of Sec. \ref{theclean} in this regime the slope of $S$ 
directly provides  the roughening
exponent $\zeta$ we are interested in. However, we have to face the problem
that for small $q$ this condition could not be satisfied. 

\begin{figure}[!t]
\begin{center}
\includegraphics[width=0.95\columnwidth]{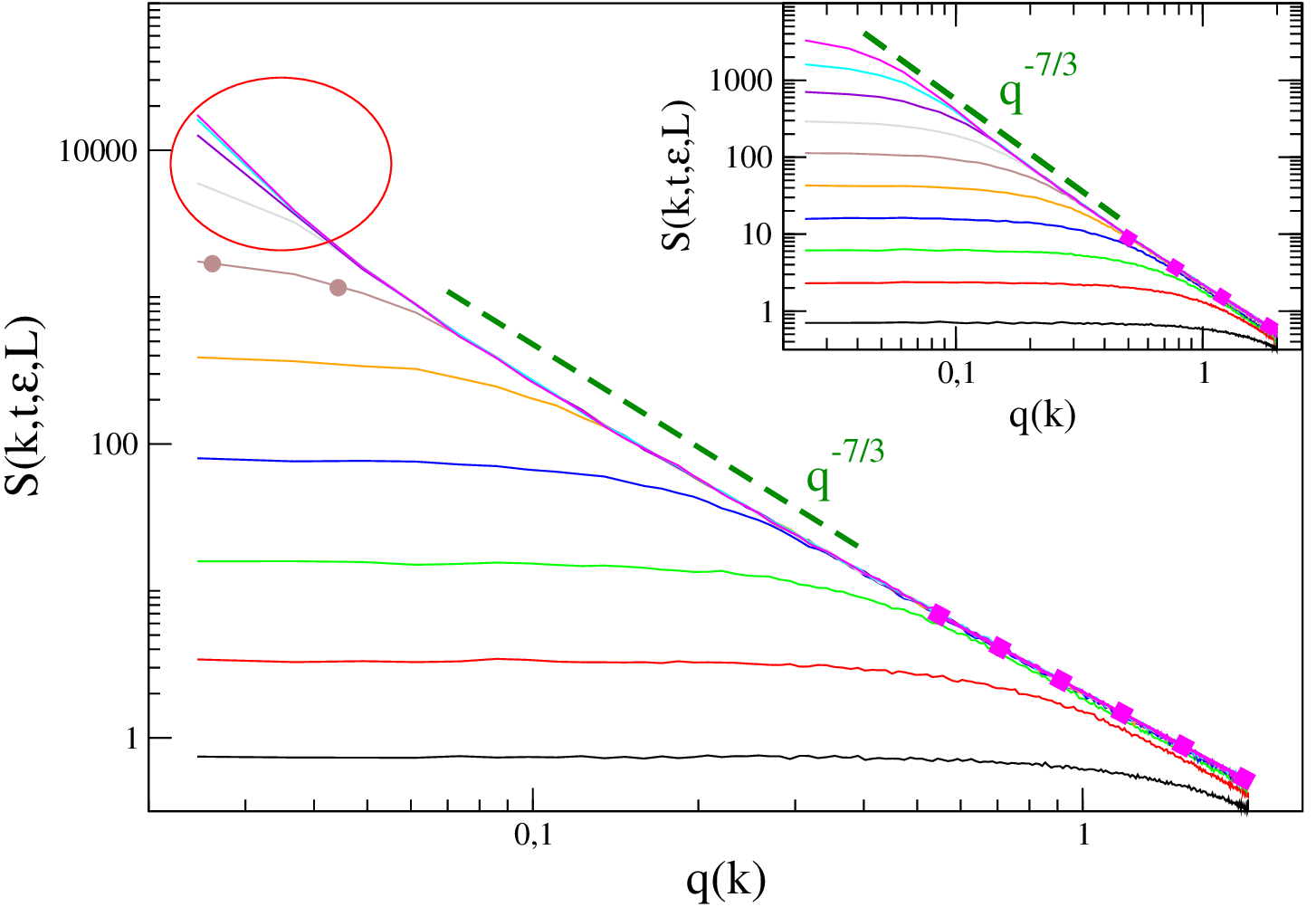}
\end{center}
\caption{$S(k,t,\epsilon=0,L)$ is plotted against $q$ for the RBIM with 
$\epsilon =0.3$ and, in the inset, for $\epsilon =1$.
Different curves correspond to exponentially increasing times 
($t=1,6,36,216,1292,7743,46416,278256,1668101,10^7$ 
from bottom to top).
The bold-dotted green line is the power-law $q^{-7/3}$ expected for
an infinite system in the limit $t\to \infty$. The regions encircled by a red
ellipse is the one where finite-size effects show-up.
The heavy-dotted magenta line is the power-law
$q^{-2}$ expected for small $x$ according to Eq. (\ref{s_clean}).}
\label{fig_RB_03}
\end{figure}

In order to restrict the analysis to the 
$q$-region where condition Eq. (\ref{twocond}) holds
we proceed as for the pure case by plotting the data for different times similarly to what done 
in the main part of Fig. (\ref{fig_clean}), as it is 
illustrated in Fig. \ref{fig_RB_03} for 
the RBIM. Then, by considering for example the intermediate time 
$t=7743$ (brown with circles) in the main part of the figure 
(with $\epsilon =0.3$) and
comparing the curve with the one at the previous time
($t=1292$) we conclude that for $t=7743$ one has $\max \left[\xi ^{-1},L^{-1}\right]=\xi ^{-1}$
and the condition (\ref{twocond}) is met at least down to the wave vector
$q\simeq 0.1$. Indeed, up to this point the two
curves (for $t=7743$ and $t=1292$) superimpose. 
For very large times ($t\ge 46416$, say) 
one observes a pattern of behavior similar to the one seen for large times in the pure case,
namely an initial upward bending of the curves before flattening again. 
This is particularly evident in the
main part of the figure in a region indicated with a red ellipse. As we will discuss further
in Sec. \ref{ranbo}, the expected value of $\zeta $ for the RBIM is $\zeta =2/3$
implying $S\simeq q^{-7/3}$ for $q\gg \xi ^{- 1}$ in an infinite system.
On the contrary, we see that at least for $t\ge 46416$ the curves reach a larger
slope before flattening again.
As in the pure case this feature identify quite clearly
the fact that $\max \left[\xi ^{-1},L^{-1}\right]=L ^{-1}$,
finite size effects become relevant and condition (\ref{twocond}) is spoiled.  
Notice that this occurs around the same wave vector $q\simeq 0.05$ of the pure case.
Interestingly, by repeating the analysis for $\epsilon =1$ (data in the inset of Fig. \ref{fig_RB_03}) we find that finite size effects are much smaller with respect
to the case with $\epsilon =0.3$: Only 
for the longest time $t=10^7$ one observes a weak upward bending of the curve. 
This difference with the case $\epsilon =0.3$ occurs because the larger the disorder is,
the slower is the growth of $\xi (t)$, and finite size effects are pushed to larger times.

By repeating the procedure described above both for the RBIM and the RFIM, for any choice of $t$ and 
of the disorder strength we end up with a definition of the $q$-sector corresponding 
to the condition (\ref{twocond}). This being done, we have purged the data in order to keep only those 
in the sector where Eq. (\ref{twocond}) is fulfilled: 
these are the only ones which have been plotted in
Figs. \ref{fig_rb},\ref{fig_rf}. Then, the slope of the curves in these figures gives 
immediate access to the exponent $\zeta$.
We stress however, that the main issue considered in this paper, namely the presence
of a rich crossover structure between different roughening regimes, can be
read out already in the large-$q$ region, as we will discuss below, where 
no doubts arise on the fulfillment of the condition in Eq. (\ref{twocond}).

The properties of the structure factor and the value of the roughening exponent 
depend on the nature of quenched disordered, either
random bonds or random fields. We will then discuss these two cases separately in the
next subsections.

\subsection{Random bonds} \label{ranbo}

In the RBIM disorder is only effective on the interfacial spins. Indeed
the contribution to the energy of spins away from the interface
is independent of the sign of the bulk-phase as long as this phase is 
uniform (namely all the spins in the bulk are aligned). This is the case 
with our simulations, since bulk fluctuations
are suppressed in the $J_0\to \infty$ limit considered here. 

When disorder is effective only on the interface the roughening
exponent is known to be $\zeta =2/3$ \cite{natterman,dueterzi}.
This value
can be understood on the basis of the following simple scaling argument:
The roughness of a piece of interface of linear size 
$\xi$ is $w\sim \xi ^\zeta$. The elastic energy associated to this deformation
is $E_{el}\sim \xi \cdot (\nabla )^2$, where $\nabla \sim \frac{w}{\xi}\sim \xi ^{\zeta -1}$ 
is the (average) local slope of the interface with respect to the flat configuration. 
Hence $E_{el}\sim \xi ^{2\zeta -1}$.
Next we want to evaluate the pinning energy associated to the random bonds 
$E_{pin}=\sum _{ij\in {\cal I}} J_{ij}$,
where the sum runs over the bonds located on the actual position of the interface ${\cal I}$. 
In writing this expression we have used the fact that disorder in this 
model is only effective around the interface,
which has the dimension of a length. Making the identification of this length
with $w$ and invoking the central limit theorem we have $E_{pin}\sim \sqrt w$. 
Balancing the two energetic contributions, $E_{el}\simeq E_{pin}$
one obtains $\zeta =2/3$.

This value of the roughening exponent  implies, through Eq. (\ref{s_clean0}),
a power-law $S\sim q^{-7/3}$ for the structure factor.
This agrees with what observed in Fig.~\ref{fig_rb}.
Indeed, for sufficiently small wavevectors $q\ll \lambda ^{-1}(\epsilon)$,
one observes that the pure-like behavior is spoiled, the curves clearly bend upward and 
the slope of  the structure factor increases. 
Fitting this slope in the range $q\in [0,0.3]$ and extracting the value of 
the roughness exponent using Eq. (\ref{s_clean0}) we find 
values between $\zeta =0.60$ and $\zeta = 0.71$ 
($\zeta = 0.60,0.62,0.65,0.66,0.69,0.70,0.71,0.7$ for 
$\epsilon=0.3,0.4,0.5,0.6,0.7,0.8,0.9,1$, respectively).
These values are compatible with the expected exponent $\zeta =2/3$. 
The smooth crossover from the pure to the disordered behavior 
and the relatively small range of wavevectors where the disordered behavior is observed 
do not allow us to provide a more precise determination of the roughness
exponent and a clearcut evidence of $\zeta =2/3$, for which
we make reference to a specific literature \cite{dueterzi}. 
The aim of this paper, however, is not to push simulations to such large sizes and long times to 
further refine the actual measurement of the roughness exponent for the RBIM, 
but rather to 
describe the rich pattern of crossover induced by the presence of disorder,
that we will discuss further below. For this task the quality of the data presented here is  
by far sufficient.

\subsection{Random fields} \label{ranfi}

In the RFIM, disorder not only acts on the interface but also 
in the bulk ${\cal B}$. Indeed, given a configuration of the 
quenched random field, the energy of 
an ordered domain depends on the sign of the magnetization,
because an excess of, say, positive random fields favors spins ordering up.
In this situation, when disorder couples to the bulk, the roughening
exponent is known to be $\zeta =1$ \cite{natterman,nota}.
Also this value can be understood in terms of the simple scaling argument
presented in Sec. \ref{ranbo}. For the RFIM one has 
$E_{pin}\sim E_{\cal I}+E_{\cal B}$, where the first term $E_{\cal I}=-\sum _{i\in {\cal I}} H_iS_i$
is the pinning energy due to the
sites around the interface whereas the second one $E_{\cal B}=-\sum _{i\not \in {\cal I}} H_iS_i$
is the pinning energy due to the bulk. 
In the thermodynamic limit $E_{\cal B}$, which takes contributions 
over a surface of order $\xi ^2$, dominates. Invoking again the central limit theorem and balancing the
elastic and the pinning energies one finds $\zeta =1$.

The data of Fig.~\ref{fig_rf} when the strength of the random field is in the
range $\epsilon \lesssim 0.2$ show that the slope of the structure factor, starting
from the pure-like behavior $S\sim q^{-2}$ for large $q$, increases
up to a value which is consistent with $S(k,t,\epsilon,L)\sim q^{-3}$ 
(shown with a dot-dashed indigo line) for sufficiently
small wavevectors. Recalling Eq. (\ref{s_clean0}) this implies that $\zeta$ is
consistent with the expected value $\zeta =1$. Indeed, by fitting the slope of 
$S$ over the last four points and extracting $\zeta $ we find $\zeta=0.93,1.01,1.05$
for $\epsilon =0.05,0.1,0.2$, respectively.

Let us now move the attention to the large values of $\epsilon $. Starting our
analysis from the cases $\epsilon \ge1$ one can notice that, as already remarked in 
Sec. \ref{thedisorder}, the pure-like behavior is never observed. Indeed already
at the largest wavevectors the slope of the structure factor is 
much larger than that $S\sim q^{-2}$ of a pure system.
This is better shown in the inset of Fig. \ref{fig_rf} where the effective value of $\zeta$
obtained by fitting the slope of $S$ in the range $q\ge 1$ 
is shown for any value of $\epsilon$. One sees that
$\zeta $ saturates for large $\epsilon $ to a value very well consistent
with  $\zeta =2/3$ (specifically one has $\zeta =0.63$,
$\zeta =0.67$, $\zeta =0.66$ for $\epsilon =1,2,3$ respectively), namely the 
value expected for a system -- like the RBIM -- where disorder
acts only on the interface. This unexpected fact, which represents the main
new result of this paper,  will be discussed and interpreted later.

Starting from this large-$q$ value, the slope of $S$ (still considering data for 
$\epsilon \ge 1$) gradually increases as $q$ lowers.
For instance, upon fitting the four points with the smallest $q$'s for $\epsilon =1$ 
one obtains a value yielding an (effective)
roughness $\zeta = 0.72$. This value,however, is not settled and keeps increasing,
consistently with the expectation that $\zeta =1 $ is the true asymptotic behavior.
With $\epsilon \ge 1$ this cannot be checked further because the curves end around $q\simeq 0.2$,
but, for smaller $\epsilon$, the data extend to smaller values of $q$ and 
a fit over the four smallest $q$ points for $\epsilon =0.8,0.6,0.4,0.3$ yields 
$\zeta =0.72,0.81,0.93,0.94,0.98$ respectively. 
This confirms that the true asymptotic regime with $\zeta =1$ is always
approached, for any $\epsilon$, if sufficiently small wavevectors are considered.

The pattern of behavior observed in Fig. \ref{fig_rf}
can be interpreted as follows: in the RFIM the disorder acts both on the interface
and in the bulk. The latter is bound to prevail in the limit of large
lenghscales and rules the behavior of $S$ at small $q$. However, on small distances, 
the interface contribution can compete with the bulk one
affecting the large-$q$ behavior of $S$. This would explain why the value 
$\zeta =2/3$ is observed in Fig. \ref{fig_rf} for large $q$ when $\epsilon$ is large.

\begin{figure}[!t]
\begin{center}
\includegraphics[width=0.95\columnwidth]{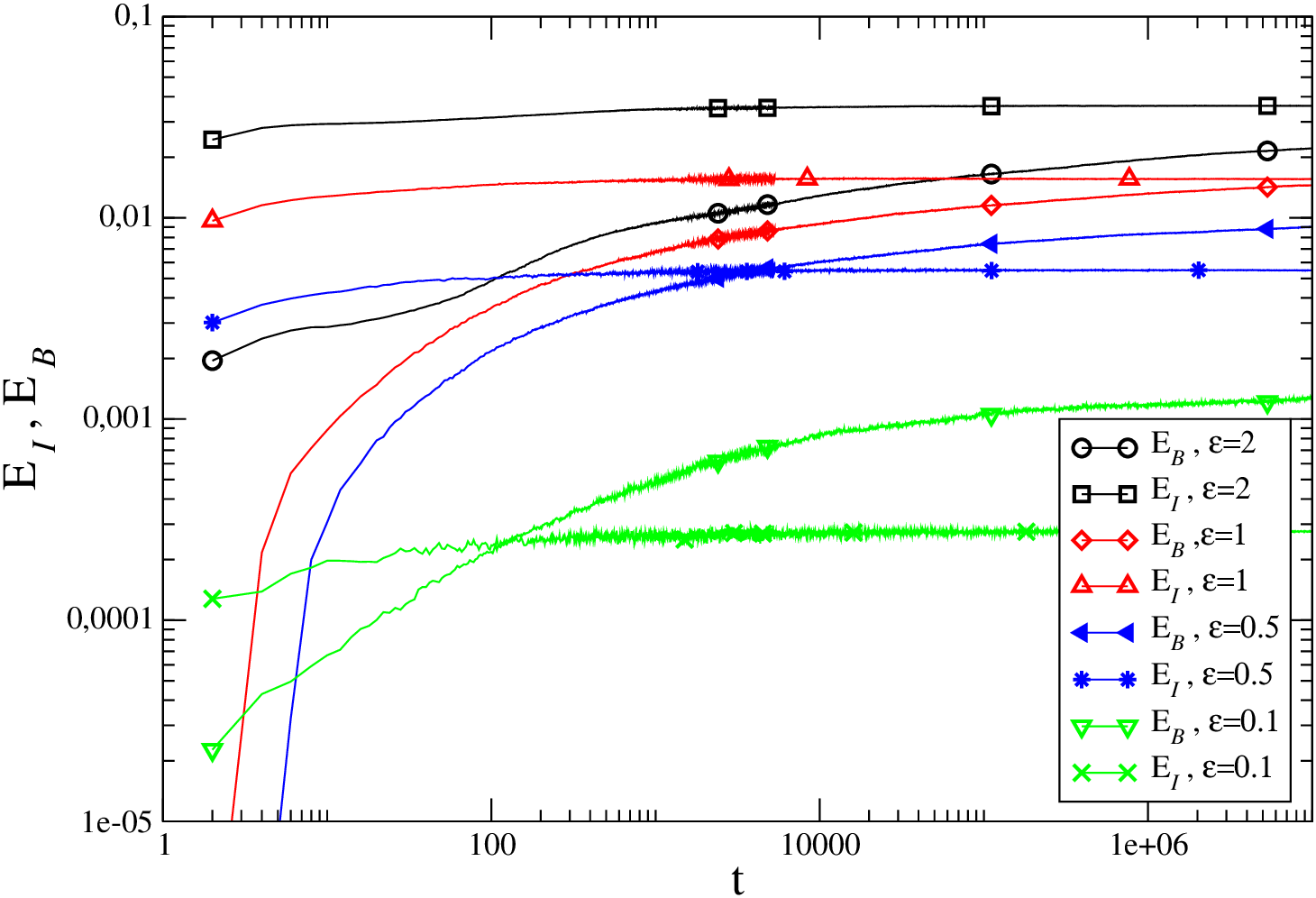}
\end{center}
\caption{$E_{\cal I}$ and $E_{\cal B}$ are plotted as a function of time 
for the RFIM and different values of $\epsilon $ (see key).}
\label{fig_compare}
\end{figure}

In order to substantiate this conjecture,
we have separately computed the energy 
$E_{\cal I}(t)$ and $E_{\cal B}(t)$ associated to the
random fields acting on the surface ${\cal I}$ or in the bulk ${\cal B}$. 
As it can be seen in Fig.~\ref{fig_compare}, $E_{\cal I}$
dominates over $E_{\cal B}$ for a certain time,
before being overtaken by the latter at the crossing time $t^*$, when
$E_{\cal I}(t^*)\simeq E_{\cal B}(t^*)$. 
Notice that $t^*$ increases very fast with the strength of the disorder:
for $\epsilon =0.1$ the bulk contribution overwhelms the surface one very soon
($t^*\simeq 200$) but for $\epsilon =1$ it is already of the order of the largest simulated time 
$t=10^7$ and goes well beyond this value for $\epsilon =2$.

One can argue that the behavior of
global quantities -- namely those which are obtained by summing up the contributions from 
all wavevectors, such as as the total energy of the system
or the roughness $W$ 
-- will resemble that of a system where disorder only acts on the
interfaces, as the RBIM, for $t\ll t^*$. At $t\simeq t^*$ a crossover occurs and,
for $t\gg t^*$ the expected asymptotic behavior of a system where
the bulk contribution prevails, as the RFIM, is observed.
Associating a length
$\Lambda (\epsilon)=\xi (t^*)$ to the crossing time $t^*$ one can say,
equivalently, that  the crossover occurs when
$\xi (t)\sim \Lambda (\epsilon)$. 

Concerning non-global quantities that depend on $q$, as the structure factor, 
for a given time $t$ one expects to observe the crossover around
$q\simeq \Lambda ^{-1}$: for $q\gg \Lambda ^{-1}$ the typical behavior
of a system where disorder only acts on the interface is observed, while the
bulk contribution 
will be manifested for $q\ll \Lambda ^{-1}$. This explains the crossover from
$S\propto q^{-7/3}$ to $S\propto q^{-3}$ observed in Fig. \ref{fig_rf}
for large $\epsilon $. 
Notice that from the data of Fig. \ref{fig_compare} one 
realizes that for $\epsilon =1$ one has $t^* \simeq 10^7$. Therefore the structure factor
with $\epsilon =1$ in  Fig. \ref{fig_rf}, which is computed at $t=10^7$ cannot
access the region where disorder in the bulk dominates. Indeed the exponent
$\zeta =1$ is not observed at any $q$.

It must be stressed that, for large $\epsilon$, a value of $\zeta $ compatible with
$\zeta =2/3$ was observed in the region of large wavevectors where any 
problem related to purging the data from finite-size/finite-times effects is
avoided and the analysis is more reliable. Not only this value,
but the whole crossover leading from the pure like behavior
to the one of a system with disorder acting only around the
interfaces is displayed in this $q$-region. Indeed, by fitting $\zeta $ from the
curves for all values of $\epsilon$ in the range $q\ge 1$ we get a typical 
sigmoidal behavior starting from $\zeta =1/2$ and converging to $\zeta \simeq 2/3$,
as it is shown in the inset of Fig. \ref{fig_rf} 

The crossover structure discussed insofar, 
where in a certain range a system with bulk disorder
behaves as one where randomness acts only on the domain wall, is the main new
result of this paper.
Let us stress that ignoring this structure
can lead to a wrong determination
of the true roughening exponent. For instance, fitting naively the structure factor 
of Fig. \ref{fig_rf} with $\epsilon =1$ could lead
to the wrong conclusion that the asymptotic roughening exponent for the RFIM is $\zeta =2/3$ whereas this is only a manifestation of a (long-lasting) preasymptotic behavior associated to the effect of disorder on the interface.
Pushing the simulations in order to access
smaller wavevectors, one could see the crossover towards the truly asymptotic behavior
with $\zeta =1$. 

Finally, we remark that the results of this paper can be fully accounted for
in simple terms as due to a crossover between the behavior of the system with
surface and bulk disorder and does not seem to be related to a so-called 
anomalous scaling behavior \cite{altraref}.  

\section{Conclusions} \label{theconclusion}

In this paper we have studied the roughening properties of an interface in
a disordered medium by computing the dynamical structure factor $S$
in the two-dimensional RBIM and RFIM.
The main difference between the two
is represented by the fact that the action of the disorder in the former is
limited to the interface while in the second it acts also in the bulk.
Accordingly, their interfaces belong to different universality classes characterized  
by the roughening exponents $\zeta =2/3$ and $\zeta =1$,
respectively.

Our simulations show a rich behavior of $S$ which can be interpreted as due
to crossover phenomena. For both kinds of disorder a first crossover
is observed around a certain wavevector $q\simeq \lambda ^{-1}(\epsilon )$,
where $\lambda (\epsilon )$ is a typical size below which disorder is ineffective,
between the behavior of a pure system -- for $q>\lambda ^{-1}$ -- and that of a 
disordered one for $q< \lambda ^{-1}$. In the region where disorder is active 
one recovers the expected roughness exponent $\zeta =2/3$ when the disorder
acts only on the interfaces, as for the RBIM.
The behavior is richer when disorder acts also in the bulk, as in the RFIM, with
the presence of an additional crossover at a certain wavevector 
$q\simeq \Lambda ^{-1}(\epsilon)<\lambda ^{-1}(\epsilon)$ between a region -- for $\lambda ^{-1}<q<\Lambda ^{-1}$
-- where the effects of the disorder on the interface prevails, yielding an (effective)
exponent $\zeta =2/3$, and the small-$q$ sector where bulk effects are important
and the (true) roughening exponent $\zeta =1 $ is observed. For this reason the 
numerical determination
of the asymptotic value of $\zeta $ can be, depending on the values of $\lambda$ 
and $\Lambda$,
hindered by the preasymptotic surface effects.

The pattern of crossovers observed in this paper for the RBIM and the RFIM are
expected to be representative of a large class of systems where disorder acts
prevalently on the interfaces or in the bulk as well.

\vspace{1cm}

{\bf Acknowledgments}
F.Corberi acknowledges financial support by MURST PRIN 2010HXAW77\_005.

We warmly thank A.B. Kolton for discussions and suggestions.

\appendix

\section{Discussion of the finite-size effects}

\begin{figure}[!t]
\begin{center}
\includegraphics[width=0.65\columnwidth]{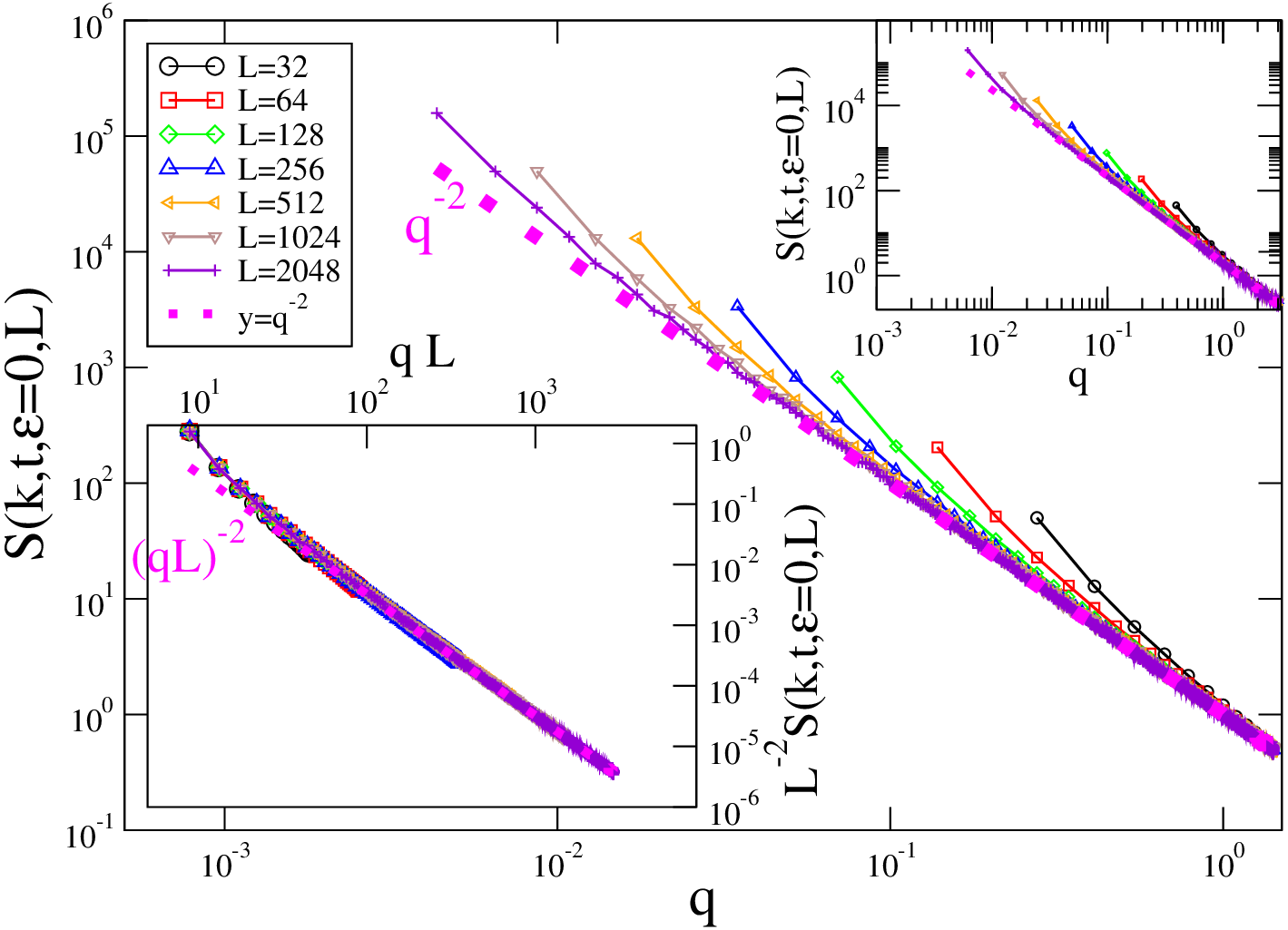}
\end{center}
\caption{$S(k,t,\epsilon =0,L)$ is plotted against $q$ for the pure case, at a very large 
time $t=10^6$, for different system syzes $L$ (see key). The bold-dotted magenta line is the
power-law expected for an infinite system in the limit $t\to \infty$.
In the upper inset the same plot is made for a system with a different geometry (see text).
In the lower inset the data of the main figure are collapsed 
by plotting $L^{-2}S(k,t,\epsilon =0,L)$ against $qL$.}
\label{fs_eps0}
\end{figure}

In this section we present a study of the finite-size effects
observed in the model discussed in the paper. Fig. \ref{fs_eps0} shows the behavior
of the structure factor for the pure system of different sizes $L$ 
at a very large time $t=10^6$.
For the pure case such time is well beyond the equilibration time of the interface,
therefore by increasing further $t$ one shouldn't see any modification to 
the pattern of Fig. \ref{fs_eps0}. In this figure one observes that the curves
follow the behavior $q^{-2}$ expected for an infinite system down to a certain
$L$-dependent wavevector below which they bend upward, as already discussed in the
paper. The wavevector where this happens decreases as $L^{-1}$ when the system size increases,
signalling that this effect is due to the finite-size of the system.
A more quantitative test can be done by trying to rescale the data, 
according to Eq. (\ref{s_clean2}), by plotting $L^{-2}S$ against $qL$.
This results in an excellent data collapse, confirming the finite-size nature
of the upward bending of the curves. Notice that, as it is
shown in the upper inset of Fig. \ref{fs_eps0}, the same phenomenon is observed
also by using the different geometry considered in \cite{sugref1}. 
This amounts to tilt the lattice by $45^o$, seeding the initial interface vertically in the 
middle of the system and applying periodic and antiperiodic boundary conditions 
in the horizontal and vertical directions, respectively. The data in the inset are characterized
by the same upward-bending finite-size effect of the main part of the figure, signaling
that this is not peculiar to our choice of the boundary conditions.
Furthermore, we mention the fact that a similar upward bending due to 
finite size effects is also observed, in the different context of
unisotropic roughening in 2+1 dimension, in \cite{vivo}.

\begin{figure}[!t]
\begin{center}
\includegraphics[width=0.45\columnwidth]{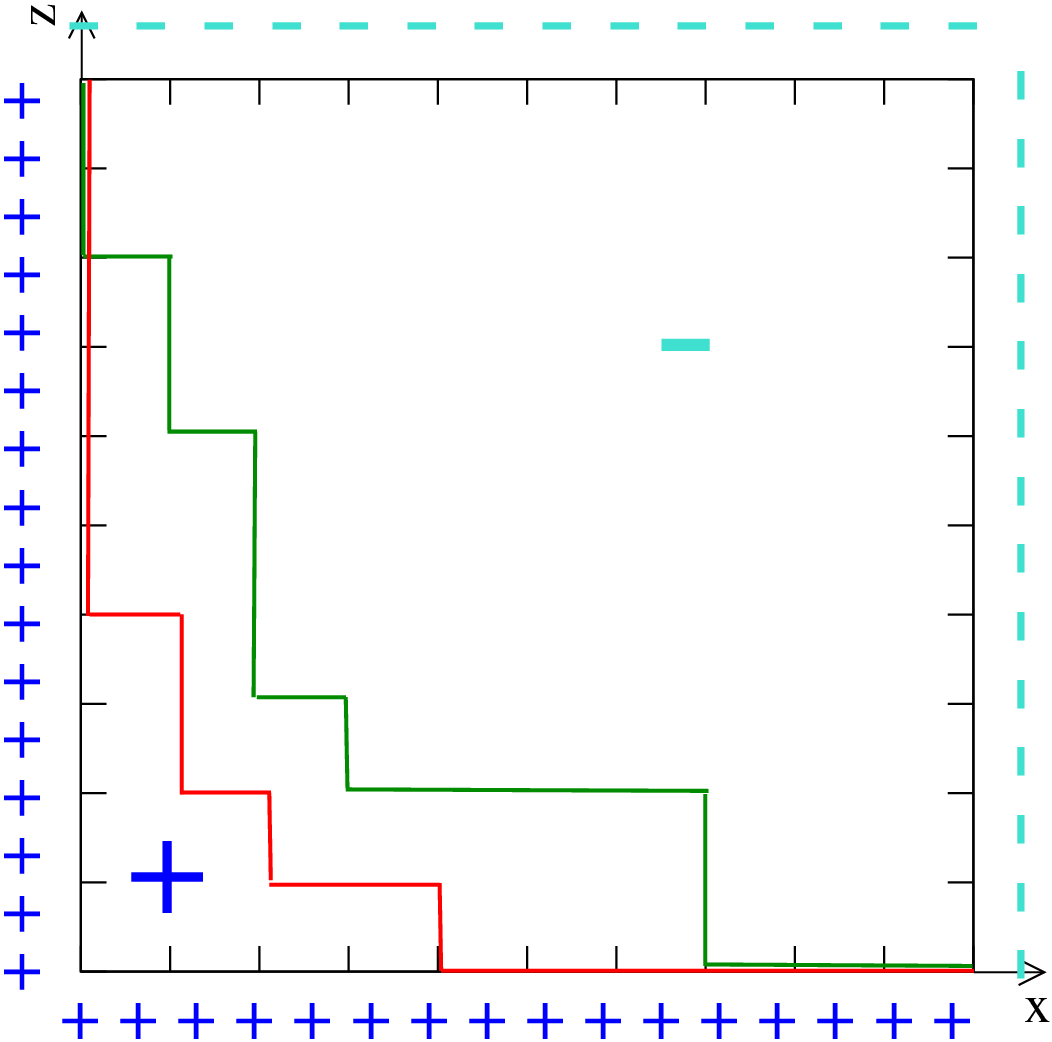}
\hspace{1cm}
\includegraphics[width=0.45\columnwidth]{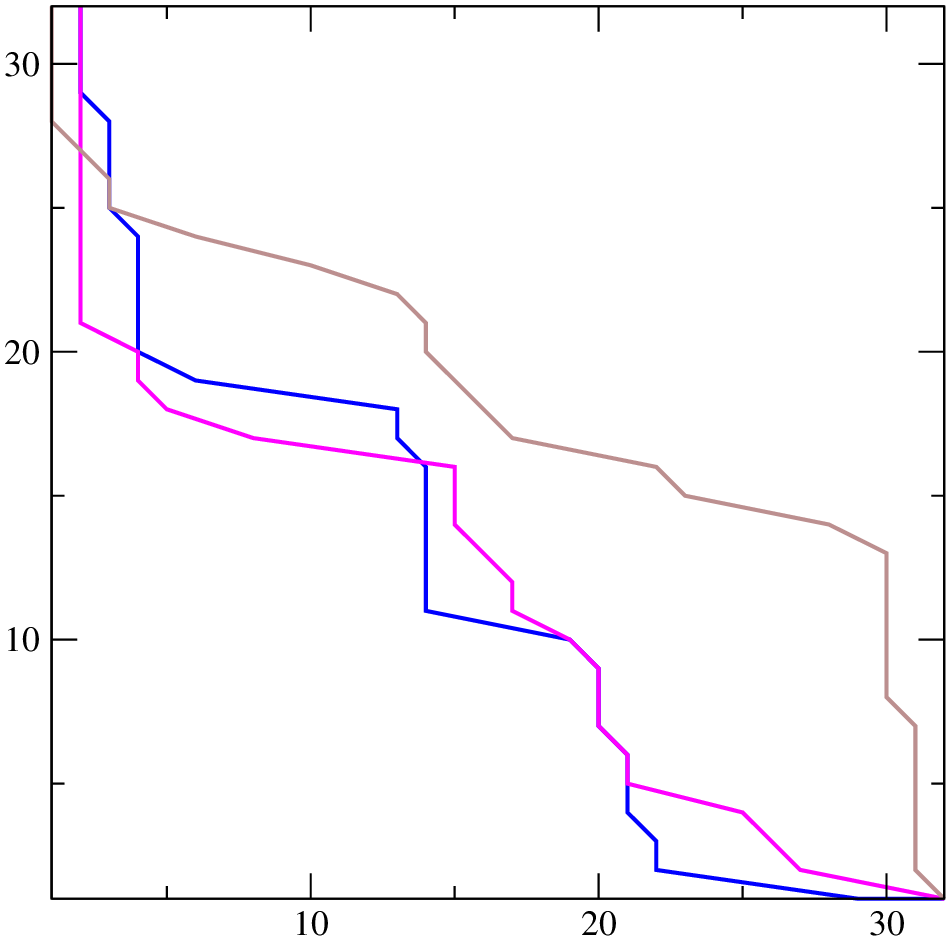}
\end{center}
\caption{Left panel. The position of the interface at time $t=0$ (blue curve on the left)
and its evolution at a later time $t$ (red curve on the right) are pictorially sketched.
Right panel: real configurations of the model interface in a pure system of size $L=32$ at large times.}
\label{fin_size_sketch}
\end{figure}

It must be noticed that a similar feature is not observed in continuum theories 
for interface roughening \cite{upton} where, instead, the curves for finite $L$ depart from the
$k^{-2}$ behavior from below flattening to a finite value. This is due to the different 
way in which the finite-size is incorporated in the discrete models considered insofar. 
One can be understand this with the help of Fig. \ref{fin_size_sketch}, where in the left
panel two
possible configurations of the interface at the same large time are pictorially sketched.
Spins on the left of the interface are down and the other are up; the 
boundary conditions adopted are also represented. 
The roughness of the rightmost red interface, denoted hereafter by ${\cal R}$,
is somewhat larger than the one of the leftmost green
configuration ${\cal G}$. Due to the proximity to the boundary, ${\cal R}$ is characterized
by long flat parts, longer than in ${\cal G}$. 
Such straight segments are actually observed in the real large
time dynamics of the model interface as it is shown in the right panel. 
We can say, therefore, that configurations of the interface with larger roughness
are associated to an excess of flat parts due to the boundary condition.
With our spin-flip kinetic rule the time needed to change a configuration
is proportional to the number of spins that can be updated, namely those on
the corners of the interface. An abundance of straight parts reduces the number of
such updates and makes the corresponding configuration more long-lived with respect
to the ones with many corners. 
As a consequence, configurations reaching the border exhibit a larger $W$ 
and, at the same time, they are flatter and live longer, 
thus providing an extra contribution to the structure factor 
on the corresponding
(small) wavevectors. This explains why, in the presence of finite-size effects, the
structure factor exceeds the infinite-size behavior ($S\propto q^{-2}$) and the curves
for $S$ bend upward at small wavevectors.

\begin{thebibliography}{}
\bibitem{nota}
In ref. \cite{seppala} a value of $\zeta $ larger than $\zeta=1$ was also reported.
\end{thebibliography}
\end{document}